\documentclass[twocolumn,showpacs,preprintnumbers,amsmath,amssymb]{revtex4}
\usepackage{color}
\usepackage{pifont}
\usepackage{bbding}
\usepackage{txfonts}
\usepackage{bbm}
\usepackage{mathrsfs}
\usepackage{amsfonts}
\usepackage{longtable}
\usepackage{tipa}
\usepackage{graphicx}
\usepackage{graphics}
\usepackage{multirow}
\usepackage{dcolumn}
\usepackage{bm}
\usepackage{float}
\usepackage{natbib}
\usepackage{enumerate}

\begin{document}

\preprint{AHEP(Hadron)/AYNU[2013]}

\title{$X(1870)$ and $\eta_2(1870)$: Which can be assigned as a hybrid state?}% Force line breaks with \\

\author{Bing Chen\footnote{Corresponding author: chenbing@shu.edu.cn}}

\affiliation{%
School of Physics and Electrical Engineering, Anyang Normal
University, Anyang 455000, China
}%

\author{Ke-Wei Wei}%

\affiliation{%
School of Physics and Electrical Engineering, Anyang Normal
University, Anyang 455000, China
}%

\author{Ailin Zhang}

\affiliation{
Department of Physics, Shanghai University, Shanghai 200444, China% with \\
}%

\date{\today}% It is always \today, today,

\begin{abstract}
The mass spectrum and strong decays of the $X(1870)$ and
$\eta_2(1870)$ are analyzed. Our results indicate that $X(1870)$ and
$\eta_2(1870)$ are the two different resonances. The narrower
$X(1870)$ seems likely a good hybrid candidate. We support the
$\eta_2(1870)$ as the $\eta_2(2^1D_2)$ quarkonium. We suggest to
search the isospin partner of $X(1870)$ in the channels of
$J/\psi\rightarrow\rho f_0(980)\pi$ and $J/\psi\rightarrow\rho
b_1(1235)\pi$ in the future. The latter channel is very important
for testing the hybrid scenario.
\end{abstract}

\maketitle

\section{\label{sec:level1}INTRODUCTION}

A isoscalar resonant structure of $X(1870)$ was observed by the
BESIII Collaboration with a statistical significance of 7.2$\sigma$
in the processes $J/\psi\rightarrow\omega
X(1870)\rightarrow\omega\eta\pi^+\pi^-$ recently~\cite{Bes6}. Its
mass and width were given as
\begin{center}
$M=1877.3\pm6.3^{+3.4}_{-7.4}$MeV,~~~~~~
$\Gamma=57\pm12^{+19}_{-4}$MeV.
\end{center}
Here the first errors are statistical and the second ones are
systematic. The product branching fraction of $\mathcal
{B}(J/\psi\rightarrow\omega X(1870))\cdot\mathcal
{B}(X(1870)\rightarrow a^{\pm}_0(980)\pi^\mp)\cdot\mathcal
{B}(a^{\pm}_0(980)\rightarrow\eta\pi^\pm)=[1.5\pm0.26(stat)^{+0.72}_{-0.36}(syst)]\times10^{-4}$
was also presented~\cite{Bes6}. But the quantum numbers of $X(1870)$
are still unknown, then the partial wave analysis is required in
future.

The mass of $X(1870)$ is consistent with the $\eta_2(1870)$, but the
width is much narrower than the $\eta_2(1870)$. In the tables of the
Particle Data Group (PDG)~\cite{PDG}, the available mass and width
of $\eta_2(1870)$ are
\begin{center}
\emph{M} = 1842 $\pm$ 8MeV,~~~~~~~~~~~$\Gamma$ = 225 $\pm$ 14MeV.
\end{center}

\begin{figure}[ht]
\begin{center}
\includegraphics[width=8.4cm,keepaspectratio]{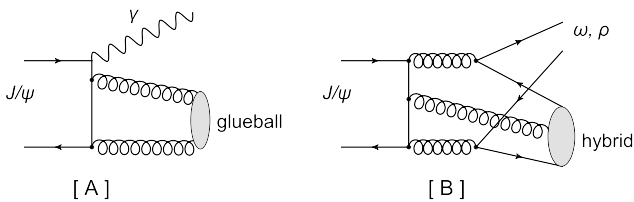}
\caption{[A]. A prior production of glueball in the $J/\psi
\rightarrow \gamma X_G$; [B]. A prior production of hybrid in the
$J/\psi \rightarrow \omega X_H$.}\label{fig1}
\end{center}
\end{figure}

The $\eta_2(1870)$ has been observed in $\gamma\gamma$
reactions~\cite{Crystall,DESY}, $p\bar{p}$
annihilation~\cite{pp1,pp2,WA1021,WA1022} and radiative $J/\psi$
decays~\cite{Bes7}. It should be stressed that radiative $J/\psi$
decay channels (Fig.\ref{fig1}[A]) and $p\bar{p}$ annihilation
prosesses are the ideal glueball hunting grounds. But the glueball
production is suppressed in $\gamma\gamma$ reaction. By contrast,
the hadronic $J/\psi$ decay are considered ``hybrid rich''
(Fig.\ref{fig1}[B]).

Furthermore, the  branching ratio $\mathcal
{R}_1=\frac{\Gamma(\eta_2(1870)\rightarrow
a_2(1320)\pi)}{\Gamma(\eta_2(1870)\rightarrow
a_0(980)\pi)}=32.6\pm12.6$ reported by the WA102 Collaboration
indicates that the decay channel of $a_0(980)\pi$ is tiny for
$\eta_2(1870)$~\cite{WA1021}. This has been confirmed by an
extensive re-analysis of the Crystal Barrel data~\cite{Bugg1}.
Differently, the analysis of BESIII Collaboration indicates that the
$X(1870)$ primarily decay via the $a_0(980)\pi$ channel~\cite{Bes6}.
Then the present measurements of the decay widths, productions, and
decay properties suggest that $\eta_2(1870)$ and $X(1870)$ are two
different isoscalar mesons.

If the production process $J/\psi\rightarrow\omega X(1870)$ is
mainly hadronic, the quantum numbers of $X(1870)$ should be
$0^+0^{-+}$, $0^+1^{++}$ or $0^+2^{-+}$. One notices that the
predicted masses for the light $0^+0^{-+}$, $0^+1^{++}$ and
$0^+2^{-+}$ hybrids overlap 1.8GeV in the Bag
model~\cite{Bag1,Bag2}, the flux-tube model~\cite{tube1,tube2} and
the constituent gluon model~\cite{gluon}. In addition, the decay
width of isoscalar $2^{-+}$ hybrid is expected to be
narrow~\cite{Swanson}. Therefore, $X(1870)$ becomes a possible
$2^{-+}$ hybrid candidate.

In addition, the predicted masses of $0^{-+}$ and $2^{-+}$ glueball
are much higher than 1.8 GeV by lattice gauge
theory~\cite{latt1,latt2,latt3}. Therefore, $X(1870)$ is not likely
to be a glueball state. Moreover, the molecule and fourquark states
are not expected in this region ~\cite{Swanson}. Then the unclear
structure $X(1870)$ looks more like a good hybrid candidate. But the
actual situation is much complicated because the nature of
$\eta_2(1870)$ is still ambiguous:
\begin{enumerate}[(i)]
 \item Since no evidences have been found in the decay mode of
$K\bar{K}\pi$, the $\eta_2(1870)$ disfavors the $1^1D_2$ $s\bar{s}$
quarkonium assignment. The mass of $\eta_2(1870)$ seems much smaller
for the $2^1D_2$ $n\bar{n}$ ($n\bar{n}\equiv
(u\bar{u}+d\bar{d})/\sqrt{2}$) state in the Godfrey-Isgur (GI) quark
model~\cite{Isgur}. Therefore the $\eta_2 (1870)$ has been assigned
as the $2^{-+}$ hybrid state~\cite{Bugg1,Bugg2,Klmept1,Amsler1}.
 \item  However, Li and Wang pointed out that the mass, production, total decay width, and decay
pattern of the $\eta_2(1870)$ do not appear to contradict with the
picture of it as being the conventional $2^1D_2$ $n\bar{n}$
state~\cite{Li3}.
\end{enumerate}

Therefore, systematical study of the mass spectrum and strong decay
properties is  urgently required for $X(1870)$ and $\eta_2(1870)$.
Some valuable suggestions for the experiments in future are also
needed.

The paper is organized as follows. In Sec.II, the masses of
$X(1870)$ and $\eta_2(1870)$ will be explored in the GI relativized
quark model and the Regge trajectories (RTs) framework. In Sec.III,
the decay processes that a isoscalar meson decays into light scalar
(below 1 GeV) and pseudoscalar mesons will discussed. The two-body
strong decays $X(1870)$ and $\eta_2(1870)$ will be calculated within
the $^3P_0$ model and the flux-tube model. Finally, our discussions
and conclusions will be presented in Sec.IV.

\section{Mass spectrum}
In the Godfrey-Isgur relativized potential model~\cite{Isgur}, the
Hamiltonian consists of the central potential and a kinetic term in
a ``relativized'' form
\begin{equation}
H=\sqrt{\vec{p}_q^2+m^2_q}+\sqrt{\vec{p}_{\bar{q}}^2+m^2_{\bar{q}}}+V_{q\bar{q}}(r).
\end{equation}

The funnel-shaped potentials which include a color coulomb term at
short distances and a linear scalar confining \linebreak[3] term at
large distances are usually incorporated as the zeroth-order
potential. The typical funnel-shaped potential was proposed by the
Cornell group (Cornell potential) with the form~\cite{Eichten}
\begin{equation}
V_{q\bar{q}}(r)=-\frac{4}{3}\frac{\alpha_s}{r}+\sigma r+C.
\end{equation}

The strong coupling constant $\alpha_s$, the string tension $\sigma$
and the constant \emph{C} are the model parameters which can be
fixed by the well established experimental states. The remaining
spin-dependent terms for mass shifts are usually treated as the
leading-order perturbations which include the spin-spin contact
hyperfine interaction, spin-orbit and tensor interactions and a
longer-ranged inverted spin-orbit term. They arise from one gluon
exchange (OGE) forces and the assumed Lorentz scalar confinement.
The expressions for these terms may be found in Ref.~\cite{Isgur}.

It should be pointed out that the nonperturbative contribution may
dominate for the hyperfine splitting of light mesons, which is not
like the heavy quarkonium~\cite{Badalian}. For example, the
hyperfine shift of the $h_c(1P)$ meson with respect to the center
gravity of the $\chi_c(1P)$ mesons is much small:
$M_{cog}(\chi_c)-M(h_c)=-0.02\pm0.19\pm0.13$MeV~\cite{CLEO}.
However, for the light isovector mesons $a_0(1450)$, $a_1(1260)$,
$a_2(1320)$, and $b_1(1235)$, the hyperfine shift is $76.7\pm44.4$
MeV. Here the masses of $a_0$, $a_1$, $a_2$, and $b_1$ are taken
from PDG~\cite{PDG}. For the complexities of nonperturbative
interactions, then we are not going to calculate the hyperfine
splitting.

Now, the spin-averaged mass, $\bar{M}_{nl}$, of $nL$ multiplet can
be obtained by solving the spinless Salpeter equation
\begin{equation}
[\sqrt{\vec{p}_q^2+m^2_q}+\sqrt{\vec{p}_{\bar{q}}^2+m^2_{\bar{q}}}+V_{q\bar{q}}(r)]\psi(r)=E\psi(r).
\end{equation}

Here we employ a variational approach described in Ref.~\cite{var}
to solve the Eq.(3). This variational approach has been applied well
in solving the Salpeter equation for $c\bar{s}$~\cite{cs},
$c\bar{c}$ and $b\bar{b}$~\cite{cc} mass spectrum.

In the calculations, the basic simple harmonic oscillator (SHO)
functions are taken as the trial wave functions. It is given by
\begin{equation}
\psi_{nl}(r,\beta)=\beta^{3/2}\sqrt{\frac{2(2n-1)!}{\Gamma(n+l+\frac{1}{2})}}(\beta
 r)^l e^{-\frac{\beta^2r^2}{2}}L^{l+1/2}_{n-1}(\beta^2r^2)\nonumber
\end{equation}
in the position space. Here the SHO function scale $\beta$ is the
variational parameter.

By the Fourier transform, the SHO radial wave function in the
momentum is
\begin{eqnarray}
\psi_{nl}(p,\beta)=\frac{(-1)^n}{\beta^{3/2}}\sqrt{\frac{2(2n-1)!}{\Gamma(n+l+\frac{1}{2})}}(\frac{p}{\beta})^l
e^{-\frac{p^2}{2\beta^2}}L^{l+1/2}_{n-1}(\frac{p^2}{\beta^2})\nonumber.
\end{eqnarray}

The wave functions of $\psi_{nl}(r,\beta)$ and $\psi_{nl}(p,\beta)$
meet the normalization conditions:
\begin{equation}
\int^\infty_0\psi^2_{nl}(r,\beta)r^2dr=1\nonumber;~~~~
\int^\infty_0\psi^2_{nl}(p,\beta)p^2dp=1\nonumber.
\end{equation}

In the variational approach, the corresponding $\bar{M}_{nl}$ are
given by minimizing the expectation value of $H$
\begin{equation}
\frac{d}{d\beta}E_{nl}(\beta)=0.
\end{equation}
where
\begin{equation}
E_{nl}(\beta)\equiv\langle
H\rangle_{nl}=\langle\psi_{nl}|H|\psi_{nl}\rangle.
\end{equation}

When all the parameters of the potential model are known, the values
of the harmonic oscillator parameter $\bar{\beta}$ can be fixed
directly. With the values of $\bar{\beta}$, all the spin-averaged
mass $\bar{M}_{nl}$ will be obtained easily. $\bar{M}_{nl}$ obtained
in this way trend to be better for the higher-excited
states~\cite{Roberts}.

It is unreasonable to treat the spin-spin contact hyperfine
interaction as a perturbation for the ground states, because the
mass splitting between pseudoscalar mesons and vector mesons are
much large. Then we consider the contributions of
$V_{\vec{s}\cdot\vec{s}}(r)$ for the $1S$ mesons. The following
Gaussian-smeared contact hyperfine interaction~\cite{ss} is taken
for convenience,
\begin{equation}
V_{q\bar{q}}^{\vec{s}\cdot\vec{s}}(r)=\frac{32\pi\alpha_s}{9m^2_q}(\frac{\kappa}{\sqrt{\pi}})^3e^{-\kappa^2r^2}\vec{S}_q\cdot\vec{S}_{\bar{q}}.
\end{equation}

In this work, we choose the model parameters as follows: $m_u$ =
$m_d$ = 0.220 GeV, $m_s =$ 0.428 GeV, $\alpha_s =$ 0.6, $\sigma =$
0.143 GeV$^2$, $\kappa =$ $0.37$ GeV, and $C =$ $-0.37$ GeV. We take
the smaller value of $\sigma$ here rather than the value in
Ref.~\cite{Isgur}. The smaller $\sigma$ was obtained by the relation
between the slope of the Regge trajectory for the Salpeter equation
$\alpha'$ and the slope $\alpha'_{st}$ in the string
picture~\cite{Badalian}. The Gaussian smearing parameter $\kappa$
seems a little smaller than that in Ref.~\cite{Isgur}. However, the
$\kappa$ is usually fitted by the hyperfine splitting of low-excited
$nS$ states in the literatures with a certain arbitrariness.

The values of $\bar{M}_{nL}$ and $\bar{\beta}$ for the states $2S$,
$3S$, $4S$, $1P$, $2P$, $3P$, $1D$, $2D$, $3D$, $1F$, $2F$, $1G$ and
$1H$ are listed in Table \ref{tableI}. The experimental masses for
the relative mesons are taken from PDG~\cite{PDG}.

\begin{table}[!htb]
\renewcommand\arraystretch{1.2}
\begin{ruledtabular}
\begin{tabular}{c|ccc|ccc}
States &$\bar{M}_{nl}(n\bar{n})$ &$\bar{\beta}$ &Expt.~\cite{PDG}&
$\bar{M}_{nl}(s\bar{s})$& $\bar{\beta}$ &Expt.~\cite{PDG}\\
\hline
1S& -     & 0.44~0.34 & -     &-    & 0.42~0.39 & -  \\
2S& 1.399 & 0.310     & 1.389 &1.631& 0.330 & 1.629 \\
3S& 1.859 & 0.295     &       &$\underline{2.069}$& 0.310 &   \\
4S& 2.240 & 0.290     &       &2.436& 0.300 &   \\
1P& 1.252 & 0.310     & 1.257 &1.460& 0.340 & 1.478  \\
2P& 1.711 & 0.294     &       &$\underline{1.926}$& 0.315 &   \\
3P& 2.110 & 0.290     &       &2.308& 0.300 &   \\
1D& 1.661 & 0.280     & 1.672 &$\underline{1.883}$& 0.300 &    \\
2D& $\underline{2.067}$ & 0.276     &       &2.272& 0.292 &   \\
3D& 2.417 & 0.275     &       &2.609& 0.288 &   \\
1F& 1.924 & 0.277     &       &2.128& 0.295 &   \\
2F& 2.287 & 0.275     &       &2.478& 0.290 &   \\
1G& 2.161 & 0.275     &       &2.350& 0.292 &   \\
1H& 2.377 & 0.273     &       &2.554& 0.287 &   \\
\end{tabular}
\end{ruledtabular}
\caption{\label{tableI}The spin-averaged mass (unit: GeV) and the
harmonic oscillator parameter $\bar{\beta}$ (unit: GeV$^{-1}$) of
the states $2S$, $3S$, $4S$, $1P$, $2P$, $3P$, $1D$, $2D$, $3D$,
$1F$, $2F$, $1G$, and $1H$.}
\end{table}

Obviously, the spin-averaged masses of the $2S$, $1P$, $1D$
$n\bar{n}$ and $1P$, $2S$ $s\bar{s}$ mesons are consistent with the
experimental data. Indeed, the predicted masses of higher excited
states here are also reasonable, $e.g.$, $a_4(2040)$ and $f_4(2050)$
are very possible the $F-$wave $n\bar{n}$ isovector and isoscalar
mesons with the masses of $1996^{+10}_{-9}$MeV and $2018\pm{11}$MeV,
respectively ~\cite{PDG}. The predicted spin-averaged mass of $1F$
is not incompatible with experiments. Our results are also overall
in good agreement with the expectations from Ref.~\cite{long1}. The
trend that a higher excited state corresponds to a smaller
$\bar{\beta}$ coincides with Ref.~\cite{Godfrey,Close,Li}. For
considering the spin-spin contact hyperfine interaction, there are
two $\bar{\beta}$s for the $1S$ mesons. The larger one corresponds
to the $1^1S_0$ state, the smaller one the $1^3S_1$ state.

As shown in Ref.~\cite{long1,long2}, the confinement potential
$V_{conf}(r)$ is determinant for the properties of higher excited
states. In Ref.~\cite{long1}, the masses for higher excited states
with $\sigma=0.143$GeV$^2$ and $\alpha_s=0$ are closer to
experimental data than the results given in Ref.~\cite{Isgur}. Then
we ignored the Coulomb interaction for $1D$, $2D$, $1F$, $1G$ and
$1H$ states. In this way, $\bar{M}_{nl}$ for these states increase
about 100MeV.

The masses of $\eta'(3^1S_0)$, $f'_1(2^3P_1)$, $\eta'_2(1^1D_2)$ and
$\eta_2(2^1D_2)$ are usually within $1.8 \sim 2.1$GeV in various
quark potential models~\cite{Isgur,Ebert,Sorace,Vijande} (see in
Table \ref{tableII}). The predicted spin-averaged masses of
$3S(s\bar{s})$, $2P(s\bar{s})$, $1D(s\bar{s})$ and $2D(n\bar{n})$
are also within this mass regions (see in Table \ref{tableI}). Due
to the uncertainty of the potential models, absolute deviation from
experimental data are usually about 100$\sim$150 MeV for the higher
excited states. Comparing with these predicted masses, $X(1870)$
disfavors the $\eta'(3^1S_0)$ assignment for its low mass. But the
possibilities of $f'_1(2^3P_1)$, $\eta'_2(1^1D_2)$ and
$\eta_2(2^1D_2)$  still exist. Here we don't consider the
possibility of $X(1870)$ as the $\eta(3^1S_0)$ state because
$\eta(1760)$ looks more like a good $\eta(3^1S_0)$
candidate~\cite{Li5,Liu,Yu}.

\begin{table}[!htb]
\renewcommand\arraystretch{1.2}
\begin{tabular}{p{1.28cm}| p{1.36cm}p{1.36cm}p{1.36cm}p{1.2cm}}
\hline\hline
States &$\eta'(3^1S_0)$& $ f'_1(2^3P_1)$  & $\eta'_2(1^1D_2)$ & $\eta_2(2^1D_2)$ \\
\hline
Ref.~\cite{Isgur}& $-$  & 2030 & 1890  & 2130$^\dag$\\
Ref.~\cite{Ebert}& 2085  & 2016 & 1909  & 1960\\
Ref.~\cite{Sorace}& 2099  & 1988 & 1851  & $-$\\
Ref.~\cite{Vijande}& $-$  & $-$ & 1853  & 1863\\
\hline\hline
\end{tabular}
\caption{\label{tableII}The masses predicted for $3^1S_0$($\eta'$),
$2^3P_1$($\eta'$), $1^1D_2$($\eta'$) and $2^1D_2$($\eta$) in
Refs.~\cite{Isgur,Ebert,Sorace,Vijande}.}
\end{table}

Regge trajectories (RTs) is another useful tool for studying the
mass spectrum of the light flavor mesons. In Ref.~\cite{Regge1}, the
authors fitted the RTs for all light-quark meson states listed in
the PDG tables. A global description was constructed as
\begin{equation}
M^2=1.38(4)n+1.12(4)J-1.25(4).
\end{equation}

Here, \emph{n} and \emph{J} mean the the radial and angular-momentum
quantum number. Recently, the authors of Ref.~\cite{Regge1} repeated
their fits with the subset mesons of the paper~\cite{Regge2}. They
found a little smaller averaged slopes of $\mu^2=1.28(5)$GeV$^2$ and
$\beta^2=1.09(6)$GeV$^2$, to be compared with $\mu^2=1.38(4)$GeV$^2$
and $\beta^2=1.12(4)$GeV$^2$ in the Eq.(7). Here the $\mu^2$ and
$\beta^2$ are the weighted averaged slope for radial and
angular-momentum RTs~\cite{Regge1,Regge3}.

Now $h_1(1380)$, $f_1(1420)$ and $\eta'(1475)$ have been established
as the $1^1P_1$, $1^3P_1$ and $2^1S_0$ $s\bar{s}$ states in
PDG~\cite{PDG}. With the differences between the mass squared of
$X(1870)$ and these states (Table \ref{tableIII}), $X(1870)$ could
be assigned for the $\eta'(3^1S_0)$ and $f'_1(2^1P_1)$. The mass of
$X(1870)$ is too large for the $\eta'_2(1^1D_2)$ state in the RTs.
$\eta_2(1645)$ has been assigned as the $1^1D_2$ $n\bar{n}$
meson~\cite{PDG}. Since
$M^2(X(1870))-M^2(\eta_2(1640))=0.91^{+0.04}_{-0.03}$GeV$^2$ which
is much smaller than $1.38(4)$GeV$^2$, $X(1870)$ looks unlike the
$2^1D_2$ $n\bar{n}$ state for its low mass. However, the difference
of $M^2(X(1870))-M^2(h_1(1170))=2.16^{+0.06}_{-0.05}$GeV$^2$ matches
the slopes $2.37(11)$GeV$^2$ well. Then the RTs can't exclude the
possibility of $X(1870)$ as the $2^1D_2$ $n\bar{n}$ state.

\begin{table}[!htb]
\renewcommand\arraystretch{1.2}
\begin{tabular}{p{2.0cm} p{2.0cm} p{2.0cm} p{1.58cm}}
\hline\hline
\multicolumn{4}{c}{Four possible states for \emph{X}(1870)}\\
\hline
$\eta'(3^1S_0)$ & $f_1'(2^3P_1)$  & $\eta'_2(1^1D_2)$ & $\eta_2(2^1D_2)$ \\
$\eta'(1475)$ & $f_1(1420)$  & $h_1(1380)$ & $\eta_2(1645)$ \\
$\mu^2=$1.34$^{+0.04}_{-0.03}$ & $\mu^2=$1.38$^{+0.04}_{-0.03}$  & $\beta^2=$1.60$^{+0.06}_{-0.06}$ & $\mu^2=$0.91$^{+0.04}_{-0.03}$ \\
\hline\hline
\end{tabular}
\caption{\label{tableIII}$X(1870)$ calculated in RTs for different
states are shown. The masses of $\eta'(1475)$, $f_1(1420)$,
$h_1(1380)$ and $\eta_2(1645)$ are taken from PDG~\cite{PDG}.}
\end{table}

As mentioned in the Introduction, $X(1870)$ is also a good hybrid
candidate since its mass overlaps the predictions given by different
models. The predicted masses for $0^+0^{-+}$, $0^+1^{++}$ and
$0^+2^{-+}$ $n\bar{n}g$ states by these models are collected in
Table \ref{tableIV}.

\begin{table}[!htb]
\renewcommand\arraystretch{1.1}
\begin{tabular}{p{2.88cm}| p{1.56cm}p{1.56cm}p{1.36cm}}
\hline\hline
States & $\eta_H(0^+0^{-+})$& $f_H(0^+1^{++})$  & $\eta_H(0^+2^{-+})$   \\
\hline
Bag~\cite{Bag1,Bag2}& 1.3  & heavier & 1.9  \\
Flux tube~\cite{tube1,tube2}& 1.7$\sim$1.9  & 1.7$\sim$1.9  & 1.7$\sim$1.9    \\
Constituent gluon~\cite{gluon}& 1.8$\sim$2.2   & 1.3$\sim$1.8  & 1.8$\sim$2.2   \\
\hline\hline
\end{tabular}
\caption{\label{tableIV}The masses predicted for $\eta_H(0^+0^{-+})$
$f_H(0^+1^{++})$ and $\eta_H(0^+2^{-+})$ hybrid states in
Refs.~\cite{Bag1,Bag2,tube1,tube2,gluon}.}
\end{table}

In this section, the mass of $X(1870)$ has been studied in the GI
quark potential model and the RTs framework. In the GI quark
potential model, $X(1870)$ can be interpreted as the $f'_1(2^3P_1)$,
$\eta'_2(1^1D_2)$ or $\eta_2(2^1D_2)$ state with a reasonable
uncertainty. In the RTs, $X(1870)$ favors the $\eta'(3^1S_0)$ and
$f'_1(2^3P_1)$ assignments. But the $\eta_2(2^1D_2)$ assignment
can't be excluded thoroughly. $X(1870)$ is also a good hybrid state
candidate. Since the masses of $X(1870)$ and $\eta_2(1870)$ are
nearly equal, the possible assignments of $X(1870)$  also suit
$\eta_2(1870)$. The investigations of the strong decay properties
will be more helpful to distinguish the $\eta_2(1870)$ and
$X(1870)$.

\section{The strong decay}
\subsection{The final mesons include the scalar mesons below 1 GeV}

Despite many theoretical efforts, the scalar nonet of $q\bar{q}$
mesons has never well-established. The lowest-lying scalar mesons
including $\sigma(500)$ (or $f_0(600)$), $\kappa(800)$, $a_0(980)$
and $f_0(980)$ are difficult to be described as $q\bar{q}$ states,
\emph{e.g}., $a_0(980)$ is associated with nonstrange quarks in the
$q\bar{q}$ scheme. If this is true, its high mass and decay
properties are difficult to be understood simultaneously. So
interpretations as exotic states were triggered, \emph{i.e}., as two
clusters of two quarks and two antiquarks~\cite{Maiani}, particular
quasimolecular states~\cite{molecule}, and uncorrelated four quark
states $qq\bar{q}\bar{q}$~\cite{tetraquark1,tetraquark2,tetraquark3}
have been proposed.

Though the structures of these scalar mesons below 1 GeV are still
in dispute, the viewpoint that these scalar mesons can constitute a
complete nonet states has been reached in the most literatures (as
illustrated in Fig.2). In the following, we will denote this nonet
as `` $\mathcal {S}$ '' multiplet for convenience.

\begin{figure}[ht]
\begin{center}
\includegraphics[width=6cm,keepaspectratio]{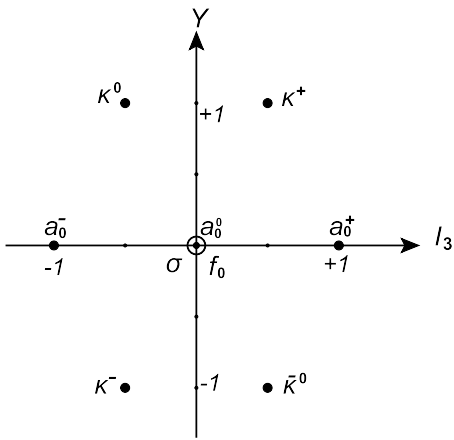}
\caption{The `` $\mathcal {S}$ '' nonet below 1 GeV shown in $Y-I_3$
plane.}
\end{center}
\end{figure}

Due to the unclear nature of the $\mathcal {S}$ mesons, it seems
much difficult to study the decay processes when the final mesons
includes a $\mathcal {S}$ member. As an approximation, $a_0(980)$,
$\sigma(500)$ and $f_0(980)$ were treated as $1^3P_0$ $q\bar{q}$
mesons in Refs.~\cite{Yu,Liu1}. In Refs.~\cite{Li3,Liu}, this kind
of decay channel was ignored. However, this kind of decay mode maybe
predominant for some mesons. For example, the observations indicate
that $f_1(1285)$, $\eta(1405)$ and $X(1870)$ primarily decay via the
$a_0(980)\pi$ channel~\cite{Bes6}.

In what follows, we will extract some useful information about this
kind of decay mode by the SU(3) flavor symmetry. We will show that
$a_0(980)\pi$, $\sigma_0\eta$ and $f_0\eta$ are the main decay
channels for the isoscalar $n\bar{n}$ and the $n\bar{n}g$ mesons
when they decay primarily through `` $\mathcal {S}$ $+$ P'' mesons,
where the sign ``P'' denotes a light pseudoscalar meson. This will
explain why $X(1870)$ has been first observed in the
$\eta\pi^+\pi^-$ channel.

We noticed that the $\mathcal {S}$ nonet could be interpreted like
the $q\bar{q}$ nonet in the diquark-antidiquark scenario. In Wilczek
and Jaffe's terminology~\cite{Jaffe,Wilczek}, the $\mathcal {S}$
mesons consist of a ``good'' diquark and a ``good'' antidiquark.
When $u$, $d$ quarks forms a ``good'' diquark, it means that the two
light quarks, \emph{u} and \emph{d}, could be treated as a
quasiparticle in color $\bar{3}$, flavor $\bar{3}$ and the spin
singlet. The ``good'' $u$, $d$ diquark is usually denoted as $[ud]$.

In the diquark-antidiquark limit, the parity of a tetraquark is
determined by $P=(-1)^{L_{12-34}}$~\cite{Santopinto} where the
$L_{12-34}$ refer to the relative angular momentum between two
clusters. Thus the $\mathcal {S}$ mesons are the lightest tetraquark
states in the diquark-antidiquark model with $L_{12-34} = 0$. The
$\mathcal {S}$ nonet in the full set of flavor representations is
\begin{eqnarray*}
(3\otimes3)_{\bar{3}}\otimes(\bar{3}\otimes\bar{3})_{3} = 8\oplus1
\end{eqnarray*}

Because the SU(3) flavor symmetry is not exact, the two physical
isoscalar mesons, $\sigma_0$ and $f_0$, are usually the mixing
states of the $|8\rangle_{I=0}$ and $|1\rangle_{I=0}$
states~\cite{Maiani},

\begin{eqnarray}
\begin{aligned}
 \left(
           \begin{array}{c}
                     f_0\\
                     \sigma_0\\
                    \end{array}
     \right)=\left(
           \begin{array}{cc}
                    \cos\vartheta  & \sin\vartheta \\
                    -\sin\vartheta  & \cos\vartheta\\
                    \end{array}
     \right)  \left(
           \begin{array}{c}
                     |8\rangle_{I=0} \\
                     |1\rangle_{I=0} \\
                    \end{array}
     \right)
\end{aligned}
\end{eqnarray}
When the mixing angle $\vartheta$ equals the so-called ideal mixing
angle, $i.e.$, $\vartheta$ = 54.74$^\circ$, the composition of the
$\sigma(500)$ and $f_0(980)$ are
\begin{eqnarray*}
\begin{aligned}
 \left(
           \begin{array}{c}
                     f_0\\
                     \sigma_0\\
                     \end{array}
     \right)=\left(
           \begin{array}{c}
                      |\frac{1}{\sqrt{2}}([su][\bar{s}\bar{u}]+[sd][\bar{s}\bar{d}])\rangle\\
                      |[ud][\bar{u}\bar{d}]\rangle\\
                    \end{array}
     \right).
\end{aligned}
\end{eqnarray*}

It seems that the deviation from the ideal mixing angle of the
$\sigma(500)$ and $f_0(980)$ is small~\cite{Maiani}. In the
following calculations, we will treat them in the ideal mixing
scheme.

Under the SU(3) flavor assumption, all the members of the octet have
the same basic coupling constant in one type of reaction, while the
singlet member have a different coupling constant. Particularly,
when a quarkonium decays into $\mathcal {S}$ and $q\bar{q}$ mesons,
there are five independent coupling constants, $i.e.$, $g_{A88}$,
$g_{A81}$, $g_{A18}$, $g_{B88}$ and $g_{B11}$, corresponding to five
different channels
\begin{center}
$\begin{cases}
\mid8\rangle_{q\bar{q}}\rightarrow\mid8\rangle_{\mathcal {S}}\otimes\mid8\rangle_{q\bar{q}}:\hspace {1cm} g_{A88}\\
\mid8\rangle_{q\bar{q}}\rightarrow\mid8\rangle_{\mathcal {S}}\otimes\mid1\rangle_{q\bar{q}}:\hspace {1cm} g_{A81}\\
\mid8\rangle_{q\bar{q}}\rightarrow\mid1\rangle_{\mathcal {S}}\otimes\mid8\rangle_{q\bar{q}}:\hspace {1cm} g_{A18}\\
\mid1\rangle_{q\bar{q}}\rightarrow\mid8\rangle_{\mathcal {S}}\otimes\mid8\rangle_{q\bar{q}}:\hspace {1cm} g_{B88}\\
\mid1\rangle_{q\bar{q}}\rightarrow\mid1\rangle_{\mathcal {S}}\otimes\mid1\rangle_{q\bar{q}}:\hspace {1cm} g_{B11}\\
\end{cases}$
\end{center}

In order to determine the relations between these coupling
constants, we shall assume the process that the $q\bar{q}$ or
$q\bar{q}g$ meson decays into a $\mathcal {S}$ and another
$q\bar{q}$ mesons obeys the OZI (Okubo-Zweig-Iizuka) rule, $i.e.$,
the two quarks in the mother meson go into two daughter mesons,
respectively. Therefore, there are four forbidden processes:
$X(\frac{1}{\sqrt{2}}(u\bar{u}-d\bar{d}))\nrightarrow a_0+s\bar{s}$,
$X(\frac{1}{\sqrt{2}}(u\bar{u}+d\bar{d}))\nrightarrow
\sigma_0+s\bar{s}$, $X(s\bar{s})\nrightarrow
f_0+\frac{1}{\sqrt{2}}(u\bar{u}+d\bar{d})$ and
$X(s\bar{s})\nrightarrow \sigma_0+s\bar{s}$. With the help of the
SU(3) Clebsch$-$Gordon coefficients~\cite{Coefficients}, the ratios
between the five coupling constants are extracted as

\begin{widetext}
\begin{equation}
\begin{aligned}
g_{A81} : g_{A18} : g_{B88} : g_{B11} : g_{A88}& =
\sqrt{2}:-\sqrt{\frac{2}{5}}(\sqrt{5}+1):-\frac{2}{\sqrt{5}}(\sqrt{5}+1):-\sqrt{\frac{2}{5}}(\sqrt{5}+1):1\\
& \approx 1.41: -2.05 : -2.89 : -2.05 : 1.00
\end{aligned}
\label{eq:wideeq}
\end{equation}
\end{widetext}

\begin{figure}[ht]
\begin{center}
\includegraphics[width=7.18cm,keepaspectratio]{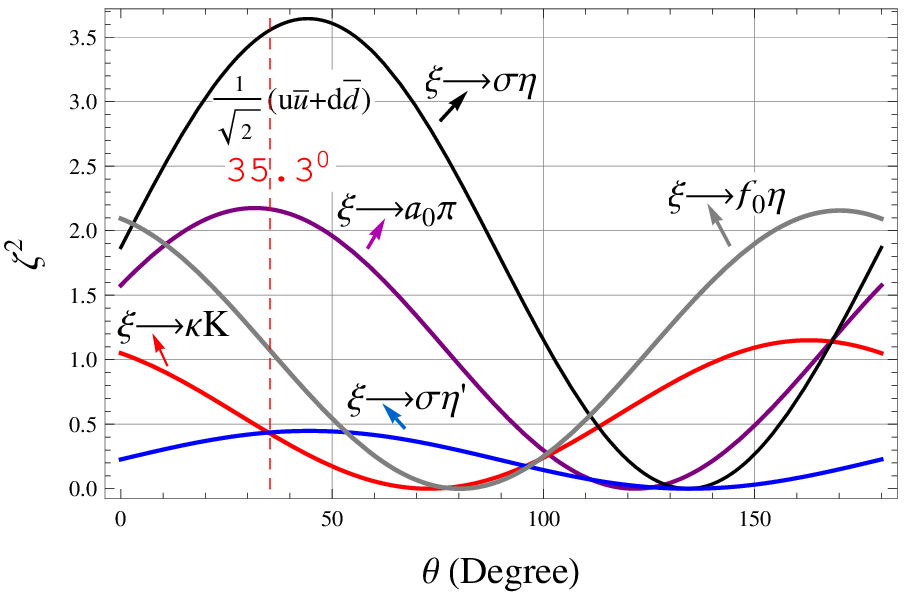}
\caption{The coefficients $\zeta^2$ of the isoscalar meson $\xi$
versus the mixing angle $\theta$.}\label{fig3}
\end{center}
\end{figure}

\begin{figure}[ht]
\begin{center}
\includegraphics[width=7.18cm,keepaspectratio]{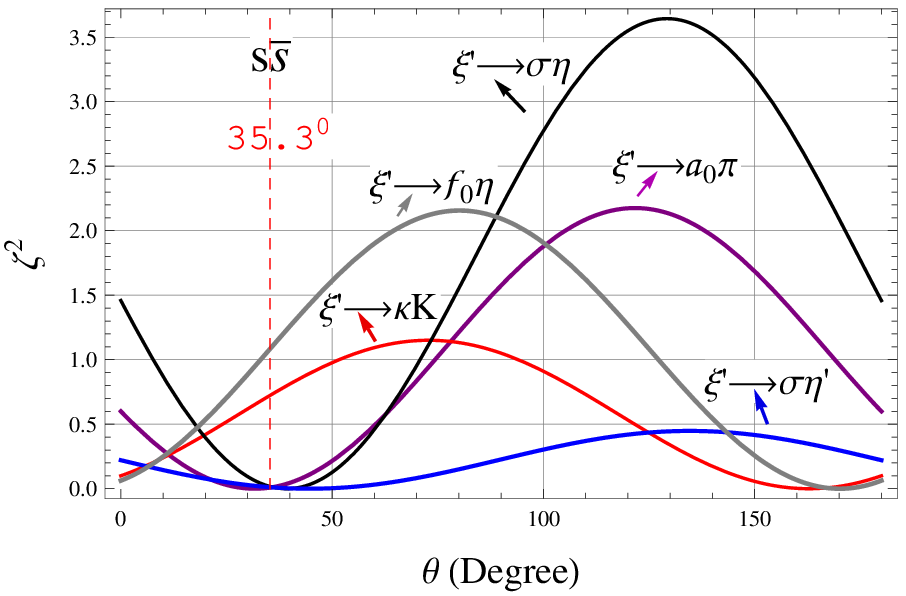}
\caption{The coefficients $\zeta^2$ of the isoscalar meson $\xi'$
versus the mixing angle $\theta$.}\label{fig4}
\end{center}
\end{figure}

It is well known that the physical states, $\eta(548)$ and
$\eta'(958)$ are the mixture of the SU(3) flavor octet and singlet.
They can be written in terms of a mixing angle, $\theta_p$, as
follows
\begin{eqnarray}
\begin{aligned}
 \left(
           \begin{array}{c}
                     \eta(548)\\
                     \eta'(958)\\
                    \end{array}
     \right)=\left(
           \begin{array}{cc}
                    \cos\theta_p  & -\sin\theta_p \\
                    \sin\theta_p  & \cos\theta_p\\
                    \end{array}
     \right)  \left(
           \begin{array}{c}
                     |8\rangle_{I=0} \\
                     |1\rangle_{I=0} \\
                    \end{array}
     \right)
\end{aligned}
\end{eqnarray}

The mixing angle $\theta_p$ has been measured by various means.
However, there is still uncertainty for $\theta_p$. An excellent fit
to the tensor meson decay widths was performed under the SU(3)
symmetry, and $\theta_p\simeq-17^o$ was obtained~\cite{Amsler1}. In
our calculation, $\theta_p$ is taken as $-17^o$. The excited
mixtures of $n\bar{n}$ and $s\bar{s}$ are denoted as

\begin{eqnarray}
\begin{aligned}
 \left(
           \begin{array}{c}
                     \xi\\
                     \xi'\\
                    \end{array}
     \right)=\left(
           \begin{array}{cc}
                    \cos\theta  & \sin\theta \\
                    -\sin\theta  & \cos\theta\\
                    \end{array}
     \right)  \left(
           \begin{array}{c}
                     |1\rangle_{I=0} \\
                     |8\rangle_{I=0} \\
                    \end{array}
     \right)
\end{aligned}
\end{eqnarray}

In this scheme, the ideal mixing occurs with the choice of
$\theta=35.3^o$. When $\xi$ and $\xi'$ decay into a $\mathcal {S}$
and pseudoscalar mesons, the relations of decay amplitudes are
governed by the coefficients $\zeta^2$ which are model-independent
in the limitation of SU(3)$_f$ symmetry. With the coupling constants
in hand, the coefficients $\zeta^2$ of $\xi$ and $\xi'$ versus the
mixing angle $\theta$ are shown in the Fig.\ref{fig3} and
Fig.\ref{fig4}. When $\xi$ and $\xi'$ occurs in the ideal mixing,
the values of $\zeta^2$ are presented in Table \ref{tableV}. In the
factorization framework, the decay difference of a hybrid and
excited $q\bar{q}$ mesons comes from the spatial
contraction~\cite{Burns}. Then the coefficients $\zeta^2$ for hybrid
states are same as these of $q\bar{q}$ quarkoniums.
\begin{table}[!htb]
\vspace{5mm}
\renewcommand\arraystretch{1.2}
\begin{tabular}{p{1.26cm}| p{1.0cm}p{1.0cm}p{1.0cm}p{1.0cm}p{0.6cm}}
\hline\hline
Decay & \multirow{2}{*}{$a_0 \pi$} & \multirow{2}{*}{$\sigma \eta$}  & \multirow{2}{*}{$\kappa K$} & \multirow{2}{*}{$f_0 \eta$} & \multirow{2}{*}{$\sigma \eta'$}  \\
channels &   &   &  &     \\
\hline
$\zeta^2[n\bar{n}(g)]$& 2.17  & 3.56 & 0.47  & 1.07 & 0.44  \\
$\zeta^2[s\bar{s}(g)]$& 0.00  & 0.00 & 0.72  & 1.08 & 0.00 \\
\hline\hline
\end{tabular}
\caption{\label{tableV}.The coefficients $\zeta^2$ of $\xi$ and
$\xi'$ in the ideal mixing.}
\end{table}

Here the mixing of $\eta(548)$ and $\eta'(958)$ has been considered.
It is sure that the $\zeta^2$ are zero for the processes, $\xi'
\rightarrow a_0 \pi$, $\xi' \rightarrow \sigma \eta$ and $\xi'
\rightarrow \sigma \eta'$, since they are OZI-forbidden. $\zeta^2$
of $\xi' \rightarrow f_0 \eta'$ hasn't been considered in Table
\ref{tableV} since $X(1870)$ lies below the threshold of $f_0
\eta'$.

As illustrated in the Fig.\ref{fig3} and Fig.\ref{fig4}, the primary
decay channels of a $s\bar{s}$ or $s\bar{s}g$ predominant excitation
are $f_0 \eta$ and $\kappa K$. If the deviation of $\theta$ from the
ideal mixing angle is not large, $X(1870)$ should be a $n\bar{n}$ or
$n\bar{n}g$ predominant state since $X(1870)$ primarily decay via
the $a_0(980)\pi$ channel. At present, only the ground $0^{-+}$ and
the $0^{++}$ isoscalar mesons deviate from the ideal mixing
distinctly. In addition, if the $X(1870)$ is produced via a diagram
of Fig.1 [B], its should also be $n\bar{n}$ or $n\bar{n}g$
predominant state.

Of course, the SU(3)$_f$ symmetry breaking will effect the ratios of
these channels listed in Table \ref{tableV}, because the
three-momentum of the these products are different. However, the
coefficients $\zeta^2$ have presented the valuable information for
these specific decay channels. When $\eta_2(1870)$ occupies the
$2^1D_2$ $n\bar{n}$ state, $X(1870)$ becomes a good $n\bar{n}g$
candidate. In the following subsection, we will explore the two-body
strong decays of $X(1870)$ within the $^3P_0$ model and the
flux-tube model. Of course, the analysis of $X(1870)$ also suit
$\eta_2(1870)$ for their nearly equal masses.

\subsection{The strong decays of $\eta_2(1870)$ and $X(1870)$}
In Ref.~\cite{Li3}, the $^3P_0$ model~\cite{Micu,Oliver1,Oliver2}
and the flux-tube model~\cite{Kokoski} were employed to study the
two-body strong decays of $\eta_2(1870)$. There, the pair production
(creation) strength $\gamma$ and the simple harmonic oscillator
(SHO) wave function scale parameter, $\beta$s, were taken as
constants.

However, a series of studies indicate that the strength $\gamma$ may
depend on both the flavor and the relative momentum of the produced
quarks~\cite{Ackleh,Bonnaz}. $\gamma$ may also depend on the reduced
mass of quark-antiquark pair of the decaying meson~\cite{Segovia}.
Firstly, the relations of the $^3P_0$ model to ``microscopic'' QCD
decay mechanisms have been studied in Ref.~\cite{Ackleh}. There, the
authors found that the constant $\gamma$ corresponds approximately
to the dimensionless combination, $\sigma/ m_q\beta$, where $m_q$ is
the mass of produced quark, $\beta$ means the meson wave function
scale and $\sigma$ is the string tension. Secondly, the momentum
dependent manner of $\gamma$ has been studied in Ref.~\cite{Bonnaz}.
It was found that $\gamma$ is dependent on the relative momentum of
the created $q\bar{q}$ pair, and the form of $\gamma(k)=A+B \exp(-C
k^2)$ with $k = |\vec{k}_3-\vec{k}_4|$ was suggested. Thirdly,
J.Segovia, \emph{et al}., proposed that $\gamma$ is a function of
the reduced mass of quark-antiquark pair of the decaying
meson~\cite{Segovia}. Based on the first and third points above,
$\gamma$ will depend on the flavors of both the decaying meson and
produced pairs. In our calculations, we will treat the $\gamma$ as a
free parameter and fix it by the well-measured partial decay widths.

In addition, the amplitudes given by the $^3P_0$ model and the
flux-tube model often contain the nodal-type Gaussian form factors
which can lead to a dynamic suppression for some channels. Then the
values of $\beta$ are important to exact the decay width for the
higher excited mesons in these two strong decay models.

In the following, the two-body strong decay of $X(1870)$ will be
investigated in the $^3P_0$ model where the strength $\gamma$ will
be extracted by fitting the experimental data. The SHO wave function
scale parameter, $\beta$s, will be borrowed from the Table
\ref{tableI} which are extracted by the GI relativized potential
model. We will also check the possibility of $X(1870)$ as a possible
hybrid state by the flux-tube model.

In the non relativistic limit, the transition operator $\mathcal
{\hat{T}}$ of the $^3P_0$ model is depicted as
\begin{widetext}
\begin{eqnarray}
\mathcal {\hat{T}}&=&-3\gamma
\sum_{\text{\emph{m}}}\langle1,m;1,-m|0,0\rangle \iint
d^3\vec{k}_3d^3\vec{k}_4\delta^3(\vec{k}_3+\vec{k}_4)\mathcal
{Y}_1^m(\frac{\vec{k}_3-\vec{k}_4}{2})\omega_0^{(3,4)}\varphi^{(3,4)}_0\chi^{(3,4)}_{1,-m}d^\dag_{3i}(\vec{k}_3)d^\dag_{4j}(\vec{k}_4)
\end{eqnarray}
\end{widetext}

Where the $\omega_0^{(3,4)}$ and $\varphi^{(3,4)}_0$ are the color
and flavor wave functions of the $q_3\bar{q}_4$ pair created from
vacuum. Thus,
$\omega_0^{(3,4)}=(R\bar{R}+G\bar{G}+B\bar{B})/\sqrt{3}$,
$\varphi^{(3,4)}_0=(u\bar{u}+d\bar{d}+s\bar{s})/\sqrt{3}$ are color
and flavor singlets. The pair is also assumed to carry the quantum
numbers of $0^{++}$, suggesting that they are in a $^3P_0$ state.
Then $\chi^{(3,4)}_{1,-m}$ represents the pair production in a spin
triplet state. The solid harmonic polynomial $\mathcal
{Y}_1^m(\vec{k})\equiv|\vec{k}|\mathcal {Y}_1^m(\theta_k,\phi_k)$
reflects the momentum-space distribution of the $q_3\bar{q}_4$.

The helicity amplitude $\mathcal {M}^{M_{J_A},M_{J_B},M_{J_C}}(p)$
of $A \rightarrow B + C$ is given by
\begin{eqnarray}\label{eq13}
\langle BC|\mathcal {\hat{T}}|A\rangle=
\delta^3(\vec{P}_A-\vec{P}_B-\vec{P}_C)\mathcal
{M}^{M_{J_A},M_{J_B},M_{J_C}}(p),
\end{eqnarray}
where \emph{p} represents the momentum of the outgoing meson in the
rest frame of the meson \emph{A}. When the mock state~\cite{Hayne}
is adopted to describe the spatial wave function of a meson, the
helicity amplitude $\mathcal {M}^{M_{J_A},M_{J_B},M_{J_C}}(p)$ can
be constructed in the $L-S$ basis easily~\cite{Oliver1,Oliver2}. The
mock state for \emph{A} meson is

\begin{equation}
\begin{aligned}
|A({n_A}&^{2S_A+1}L_A^{J_AM_{J_A}}(\vec{P}_A)\rangle\\\equiv&\sqrt{2E_A}\sum_{{M_{L_A}}{M_{S_A}}}\langle
L_AM_{L_A}S_AM_{S_A}|J_AM_{J_A}\rangle\omega_A^{12}\phi_A^{12}\chi_{S_AM_{S_A}}^{12}\\
&\times\int
d\vec{P}_A\psi_{n_A}^{L_AM_{L_A}}(\vec{k}_1,\vec{k}_2)|q_1(\vec{k}_1)q_2(\vec{k}_2)\rangle.
\end{aligned}
\end{equation}

To obtain the analytical amplitudes, the SHO wave functions are
usually employed for $\psi_{n_A}^{L_AM_{L_A}}(\vec{k}_1,\vec{k}_2)$.
For comparison with experiments, one obtains the partial decay width
$\mathcal {M}^{JL}(p)$ via the Jacob-Wick formula~\cite{Jacob}

\begin{equation}
\begin{aligned}
\mathcal {M}_{LS}(p)=&\frac{\sqrt{2L+1}}{2J_A+1}\sum_{\text{$M_{J_B}$,$M_{J_C}$}}\langle L0JM_{J_A}|J_AM_{J_A}\rangle\\
&\times\langle J_B,M_{J_B}J_C,M_{J_C}|JM_{J_A}\rangle\mathcal
{M}^{M_{J_A},M_{J_B},M_{J_C}}(p).
\end{aligned}
\end{equation}

Finally, the decay width $\Gamma(A\rightarrow BC)$ is derived
analytically in terms of the partial wave amplitudes

\begin{equation}
\begin{aligned}
\Gamma(A\rightarrow BC)=2\pi\frac{E_BE_C}{M_A}p\sum_{LS}|\mathcal
{M}_{LS}(p)|^2.
\end{aligned}
\end{equation}

More technical details of the $^3P_0$ model can be found in
Ref.~\cite{Oliver2}. The inherent uncertainties of the $^3P_0$ decay
model itself have been discussed in the
Refs.~\cite{Bonnaz,model1,model2}.

The dimensionless parameter $\gamma$ will be fixed by the 8
well-measured partial decay widths which are listed in
Table\ref{tableVI}. The $\mathcal {M}_{LS}$ amplitudes of these
decay channels are presented explicitly in the Appendix A.

\begin{table}[H]
\renewcommand\arraystretch{1.2}
\begin{tabular}{p{1.16cm} p{0.80cm}p{0.82cm}p{0.64cm}|p{1.52cm} p{0.80cm}p{0.80cm}p{0.66cm}}
\hline\hline
Decay channels & p (GeV) & $\gamma(10^3)$& $\gamma$\cite{Bonnaz} & Decay channels & p (GeV)  & $\gamma(10^3)$ &$\gamma$\cite{Bonnaz}\\
\hline
$\rho \rightarrow\pi \pi$ & 0.362  & 17.8 &9.18  & $f'_1 \rightarrow K^*\bar{K}$  & 0.158 & 4.9  & -\\
$a_2 \rightarrow \eta \pi$ & 0.535  & 11.5 & - & $f_2 \rightarrow K\bar{K}$  & 0.401& 2.9  & 6.11\\
$f_2 \rightarrow \pi \pi$ & 0.622  & 7.8 & 7.13 & $a_2 \rightarrow K\bar{K}$  & 0.434& 2.3  & 3.91\\
$\rho_3 \rightarrow \pi \pi$ & 0.833  & 4.2 & -  &$f'_2 \rightarrow K\bar{K}$  & 0.579& 2.0  & 5.66\\
\hline\hline
\end{tabular}
\caption{\label{tableVI}. Values of $\gamma$ in different channels
and comparison with the results given in Ref.\cite{Bonnaz}. Here,
$\rho(770)$, $a_2(1320)$, $f'_1(1420)$, $f_2(1270)$, $f'_2(1525)$
and $\rho_3(1690)$ have been studied.}
\end{table}

As mentioned before, $\gamma$ may depend on the flavors of both the
decaying meson and produced pairs. Then we divide the 8 decay
channels into two groups: one is $n\bar{n}\rightarrow
n\bar{n}+n\bar{n}$, the other includes $s\bar{s}\rightarrow
n\bar{s}+s\bar{n}$ and $n\bar{n}\rightarrow n\bar{s}+s\bar{n}$. The
values of $\gamma$ here are a little different from these given in
Ref.\cite{Bonnaz} where an \emph{AL}1 potential (for details of
\emph{AL}1 potential, see Ref.\cite{Roberts}) was selected to
determine the meson wave functions. Of course, the meson wave
function given by different potentials will influence the values of
$\gamma$.

It is clear in Table \ref{tableVI} that $\gamma$ decrease with
\emph{p} increase. In addition, our calculation indicates that
$\gamma$ depend on flavors of both the decaying meson and the
produced quark pairs. For example, values of $\gamma$ fixed by $a_2
\rightarrow K\bar{K}$ and $f'_2 \rightarrow K\bar{K}$ are roughly
equal.

\begin{figure}[ht]
\begin{center}
\includegraphics[width=7.18cm,keepaspectratio]{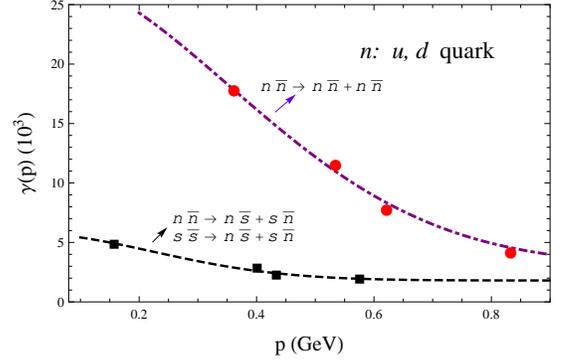}
\caption{The functions of $\gamma(p) = A + B\exp(-C p^2)$ in
different decay processes. The symbols of red ``\ding {108}'' and
black ``\ding {110}'' denote $\gamma$ values determined by the
experimental data.}\label{fig5}
\end{center}
\end{figure}

In the following calculations, we assume that the values of $\gamma$
corresponding to the processes of $s\bar{s}\rightarrow
n\bar{s}+s\bar{n}$ and $n\bar{n}\rightarrow n\bar{s}+s\bar{n}$ are
determined by one function. Similarly, we take the function,
$\gamma(p) = A + B\exp(-C p^2)$, for the creation vertex. \emph{The
function of the creation vertex here is different with the one used
in the Ref~\cite{Bonnaz}}. With the four decay channels listed in
fifth column of Table \ref{tableVI}, we fix the function as
$\gamma(p) = 1.8 + 4\exp(-10 p^2)$. For the processes of
$n\bar{n}\rightarrow n\bar{n}+n\bar{n}$ (the first column of Table
\ref{tableVI}), we fix the creation vertex function as $\gamma(p) =
3.0 + 25\exp(-4 p^2)$. The dependence of $\gamma$ on the momentum
\emph{p} are plotted in the Fig. \ref{fig5}. Obviously the functions
can describe the dependence of $\gamma$ and \emph{p} well. The
functions of creation vertex given here need further test.

Since we neglected the mass splitting within the isospin multiplet,
the partial width into the specific charge channel should be
multiplied by the flavor multiplicity factor $\mathcal {F}$ (Table
\ref{tableVII}). This $\mathcal {F}$ factor also incorporates the
statistical factor 1/2 if the final state mesons \emph{B} and
\emph{C} are identical (as illustrated in Fig.\ref{fig6}). More
details of $\mathcal {F}$ can be found in the Appendix A of
Ref.\cite{Barnes}.

\begin{figure}[ht]
\begin{center}
\includegraphics[width=7.08cm,keepaspectratio]{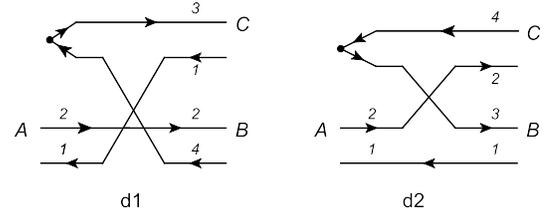}
\caption{Two topological diagrams for a $q\bar{q}$ meson decay in
the $^3P_0$ decay model. We refer to the left one as \emph{d}1 where
the produced quark goes into meson \emph{C}, and \emph{d}2 where it
goes into \emph{B}.}\label{fig6}
\end{center}
\end{figure}

\begin{table}[H]
\renewcommand\arraystretch{0.88}
\begin{center}
\begin{tabular}{p{1.48cm}|p{1.88cm}p{1.88cm}p{0.88cm}}
\hline\hline
Decay &\hspace {0.1cm}\multirow{2}{*}{$\mathcal {I}_{flavor}(d1)$}  &\multirow{2}{*}{$\mathcal {I}_{flavor}(d1)$}  &\multirow{2}{*}{$\mathcal {F}$}   \\
channels   &  &  &    \\
\hline
$\rho\rightarrow\pi\pi$     &\hspace {0.1cm}  $+1/\sqrt{2}$  &  $-1/\sqrt{2}$    & 1       \\
$f_2\rightarrow\pi\pi$   &\hspace {0.1cm} $-1/\sqrt{2}$ &  $-1/\sqrt{2}$      & 3/2     \\
$f_2\rightarrow K K$     &\hspace {0.1cm}  0      & $-1/\sqrt{2}$ & 2     \\
$f'_1\rightarrow K^*K$      &\hspace {0.1cm}  +1  &   0  & 4      \\
$f'_2\rightarrow K K$    &\hspace {0.1cm}  +1  &  0   & 2   \\
$a_2\rightarrow K K$     &\hspace {0.1cm}  0    & -1   & 1        \\
$a_2\rightarrow \eta\pi$      &\hspace {0.1cm}  $+1/2$  &   $+1/2$  & 1      \\
$\eta_2\rightarrow \omega\omega$     &\hspace {0.1cm}  $-1/\sqrt{2}$  & $-1/\sqrt{2}$    & 1/2     \\
$\eta_2\rightarrow a_i\pi$     &\hspace {0.1cm}  $-1/\sqrt{2}$  & $-1/\sqrt{2}$    & 3     \\
$\eta_2\rightarrow f_i \eta$     &\hspace {0.1cm}  $+1/2$   &   $+1/2$   & 1      \\
\hline\hline
\end{tabular}
\end{center}
\caption{\label{tableVII}The second and third columns for the flavor
weight factors corresponding to two topological diagrams shown in
Fig.\ref{fig6}. The last column for the the flavor multiplicity
factor $\mathcal {F}$. Here,
$|\eta\rangle=(|n\bar{n}\rangle-|s\bar{s}\rangle)/\sqrt{2}$ and
$|\eta'\rangle=(|n\bar{n}\rangle+|s\bar{s}\rangle)/\sqrt{2}$ have
been taken for simplicity.}
\end{table}

\begin{widetext}

\begin{table}[H]
\renewcommand\arraystretch{1.2}
\begin{center}
\begin{tabular}{p{1.52cm}|p{1.38cm}p{0.98cm}p{0.88cm}|p{0.88cm}p{1.88cm}p{0.88cm}p{2.08cm}p{0.8cm}p{1.58cm}}
\hline\hline
Decay\hspace {0.4cm}  &\hspace {0.1cm}$\eta_2(2^1D_2)$   & &    &\multicolumn{2}{l}{$\eta_H(0^+0^{-+})$}   &\multicolumn{2}{l}{$f_H(0^+1^{++})$}  &\multicolumn{2}{l}{$\eta_H(0^+2^{-+})$}\\
channels \hspace {0.4cm}  &\hspace {0.1cm} Our  & \multicolumn{2}{l|}{Ref.~\cite{Li3}}  &Our\hspace {0.4cm}  &\hspace {0.1cm}Ref.~\cite{Swanson}  & Our\hspace {0.1cm}  &\hspace {0.1cm}Ref.~\cite{Swanson}& Our\hspace {0.1cm}  &\hspace {0.1cm}Ref.~\cite{Swanson} \\
\hline
$K^* K$         &\hspace {0.1cm}  0.5  &  17.7    & 19.3    & 12.6      &\hspace {0.1cm}10\hspace {0.3cm}  5              &  4.9         & \hspace {0.1cm}24.1\hspace {0.2cm}  18.0          & 3.2      & \hspace {0.1cm}2.0\hspace {0.35cm}  1.0 \\
$\rho\rho$      &\hspace {0.1cm}  12.9 &  52.2    & 56.8    & $\times$  &\hspace {0.1cm}$\times$\hspace {0.4cm}  $\times$ &  $\times$    & \hspace {0.1cm}$\times$\hspace {0.6cm}  $\times$  & $\times$ & \hspace {0.1cm}$\times$\hspace {0.55cm}  $\times$\\
$\omega\omega$  &\hspace {0.1cm}  4.2  &  16.9    & 18.4    & $\times$  &\hspace {0.1cm}$\times$\hspace {0.4cm}  $\times$ &  $\times$    & \hspace {0.1cm}$\times$\hspace {0.6cm}  $\times$  & $\times$ & \hspace {0.1cm}$\times$\hspace {0.55cm}  $\times$ \\
$K^* K^*$       &\hspace {0.1cm}  0.2  &  2.1     & 2.3     & $\times$  &\hspace {0.1cm}$\times$\hspace {0.4cm}  $\times$ &  $\times$    & \hspace {0.1cm}$\times$\hspace {0.6cm}  $\times$  & $\times$ & \hspace {0.1cm}$\times$\hspace {0.55cm}  $\times$\\
$a_0(1450)\pi$  &\hspace {0.1cm}  16.0 &  2.4     & 2.6     & 56.3      &\hspace {0.1cm}70\hspace {0.3cm}  175            &  0.5         & \hspace {0.1cm}$\times$\hspace {0.6cm}  6                & 0.5      & \hspace {0.1cm}0.0\hspace {0.4cm}  0.6 \\
$a_1(1260)\pi$  &\hspace {0.1cm}  0.0  &  15.2    & 16.6    & $\times$  &\hspace {0.1cm}$\times$\hspace {0.4cm}  $\times$ &  57.3        & \hspace {0.1cm}14\hspace {0.5cm}  232             & $\times$ & \hspace {0.1cm}0.3\hspace {0.4cm}  $\times$  \\
$f_1(1280)\eta$ &\hspace {0.1cm}  0.0  &  0.0     & 0.0     & $\times$  &\hspace {0.1cm}$\times$\hspace {0.4cm}  $\times$ &  2.5         & \hspace {0.1cm}$-$\hspace {0.6cm}  $-$                & $\times$ & \hspace {0.1cm}0.0\hspace {0.4cm}  $\times$ \\
$a_2(1320)\pi$  &\hspace {0.1cm}  54.2 &  102.5   & 111.6   & 8.8       &\hspace {0.1cm}1\hspace {0.4cm}  16              &  35.1        & \hspace {0.1cm}5.0\hspace {0.4cm}  179.4          & 26.7     & \hspace {0.1cm}25.1\hspace {0.26cm}  67  \\
$f_2(1275)\eta$ &\hspace {0.1cm}  15.1 &  17.5    & 19.0    & 0.0       &\hspace {0.1cm}$\times$\hspace {0.4cm}  $\times$ &  1.0         & \hspace {0.1cm}$-$\hspace {0.6cm}  $-$                & 4.6      & \hspace {0.1cm}0.0 \hspace {0.36cm}  0.0 \\
\hline
$\sum\Gamma_i$  &\hspace {0.1cm}  103.3&   226.5  & 246.7   & 77.7  &\hspace {0.1cm}81\hspace {0.3cm}196  & 101.3   & 43.1\hspace {0.4cm}435.4   &35.0 &\hspace {0.1cm}27.4\hspace {0.3cm}68.6\\
Expt (MeV) & \multicolumn{3}{c|}{225$\pm$14 ~\cite{PDG}}   & & & & & \multicolumn{2}{c}{57$\pm$12$^{+19}_{-4}$ ~\cite{Bes6}} \\
\hline\hline
\end{tabular}
\end{center}
\caption{\label{tableVIII}The partial widths of $X(1870)$ and
compared with results from Refs.~\cite{Li3,Swanson}. The symbol
``$\times$'' indicates that the decay modes are forbidden and
``$-$'' denotes that the decay channels can be ignored. Here, we
collected the results given by the $^3P_0$ model from
Ref.~\cite{Li3} in the left column, the right column by the
flux-tube model. In Ref.~\cite{Swanson}, the masses are taken as 1.8
GeV for the $0^{-+}$, $1^{++}$ and $2^{-+}$ for the hybrid states.}
\label{table 3}
\end{table}
\end{widetext}

The partial decay widths of $X(1870)$ are shown in Table
\ref{tableVIII} except the channels of $\mathcal {S}$ $+$ $P$
mesons. $a_2(1320)\pi$ and $f_2(1275)\eta$ are large channels for
the $\eta_2(2^1D_2)$ $n\bar{n}$ state in our work and the
Ref.~\cite{Li3}, which are consistent with the  experimental
observations of the $\eta_2(1870)$. The partial widths of $K^* K$,
$\rho\rho$ and $\omega\omega$ are narrower in our work than the
expectations from Ref.~\cite{Li3}. $\eta_2(1870)$ has been observed
by the BES Collaboration in the radiative decay channel of
$J/\psi\rightarrow\gamma\eta\pi\pi$~\cite{Li3}. However, no apparent
$\eta_2(1870)$ signals were detected in the channels of
$J/\psi\rightarrow\gamma\rho\rho$~\cite{Mark1}
and$J/\psi\rightarrow\gamma\omega\omega$~\cite{Mark2,Bes4}.
Therefore, improved experimental measurements of the radiative
$J/\psi$ decay channels are needed  for the $\eta_2(1870)$ in
future.

\begin{figure}[ht]
\begin{center}
\includegraphics[width=4.8cm,keepaspectratio]{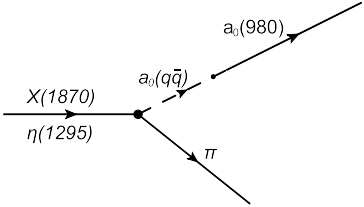}
\caption{The diagram for the `` $\mathcal {S}$ $+$ P'' channels
through a virtual intermediate $1^3P_0$ $q\bar{q}$
meson.}\label{fig7}
\end{center}
\end{figure}

Nextly, we shall evaluate the partial widths of ``$\mathcal {S}+P$''
channels which have not been listed in the table\ref{tableVIII}.
\emph{The scheme is proposed as following.} As illustrated in the
Fig.\ref{fig7}, we assume $X(1870)$ decay into $a_0(980)\pi$ via a
virtual intermediate $1^3P_0$ $q\bar{q}$ meson . We notice the
$\eta(1295)$ also dominantly decay into the $\eta \pi
\pi$~\cite{Bes6}. Its three-body decay can occur via three
intermediate processes:
$\eta(1295)\rightarrow\eta\sigma/a_0(980)\pi/\eta(\pi\pi)_{S-wave}\rightarrow\eta\pi\pi$\cite{PDG}.
With the ratio $\Gamma(a_0(980)\pi)/\Gamma(\eta\pi\pi) =
0.65\pm0.10$ and $\Gamma(\eta(1295)) = 55\pm5$MeV, the partial width
of $\eta(1295)$ decaying into $a_0(980)\pi$ is estimated no more
than 45MeV. By the $^3P_0$ model, the ratio of $\frac{\Gamma(X(1870)
\rightarrow a_0(1^3P_0)\pi)}{\Gamma(\eta(1295) \rightarrow
a_0(1^3P_0)\pi)}$ can be reached easily. If the uncertainty of the
coupling vertex of $\varepsilon(1^3P_0(q\bar{q})\rightarrow
a_0(980))$ (see in Fig.\ref{fig7}) is assumed to be canceled in the
ratio of $\frac{\Gamma(X(1870) \rightarrow
a_0(1^3P_0)\pi)\cdot\varepsilon(1^3P_0(q\bar{q})\rightarrow
a_0(980))}{\Gamma(\eta(1295) \rightarrow
a_0(1^3P_0)\pi)\cdot\varepsilon(1^3P_0(q\bar{q})\rightarrow
a_0(980))}$, the value of $\frac{\Gamma(X(1870) \rightarrow
a_0(980)\pi)}{\Gamma(\eta(1295) \rightarrow a_0(980)\pi)}$ can be
extracted roughly. Although the assumption above seems a little
rough, we just need to evaluate magnitudes of these decay channels.

$\eta(1295)$ is proposed to be the first radial excited state of
$\eta(550)$. Then the total decay widths of $\Gamma(X(1870)
\rightarrow \mathcal {S} + P)$ is evaluated no more than 12.6MeV and
$\Gamma(X(1870) \rightarrow a_0(980)\pi) \leq3.8$MeV. The BESIII
Collaboration claimed that $X(1870)$ primarily decay via
$a_0(980)\pi$~\cite{Bes6}. The small partial width of
$\Gamma(X(1870) \rightarrow a_0(980)\pi)$ also indicates that the
$X(1870)$ can't be interpreted as the $2^1D_2$ $q\bar{q}$ state.

In addition, our results do not support $X(1870)$ as the
$\eta_2(2^1D_2)$ $n\bar{n}$ state since its observed decay width is
much smaller than the theoretical estimate. The $a_2(1320)\pi$ is
the largest decay channel in our numerical results and in
Ref.~\cite{Li3} for the $\eta_2(2^1D_2)$ $n\bar{n}$ state (Table
\ref{tableVIII}). If the partial width of $a_0(980)\pi$ channel is
as large as $a_2(1320)\pi$, the predicted width of $X(1870)$ will be
much larger than the observed value.

We adopt the flux tube model to check the possibility of $X(1870)$
as a hybrid meson. The partial widths are also listed in Table
\ref{tableVIII} for the comparison. Details of the flux model are
collected in the Appendix B.

Two groups of the partial widths predicted in the
Ref.~\cite{Swanson} are quoted in the Table \ref{tableVIII}. The
left column was given by the flux tube decay model of Isgur,
Kokoski, and Paton (IKP) with the ``standard
parameters''~\cite{IKP}. The right column was by the developed flux
tube decay model of Swanson-Szczepaniak (SS). In
Ref.~\cite{Swanson}, the masses are taken as 1.8 GeV for the
$0^{-+}$, $1^{++}$ and $2^{-+}$ for the hybrid states.

For a hybrid meson, $X(1870)$ seems most possible to be the
$\eta_H(0^+2^{-+})$ state because the total widths exclude the
channels of $\mathcal {S} + P$ are much narrow in our work and in
Ref.\cite{Swanson}. It is consistent with the narrow width of
$X(1870)$.

As shown in Table \ref{tableVIII}, $X(1870)$ is impossible to be the
$\eta_H(0^+0^{-+})$ hybrid state. The predicted width in both our
work and in Ref.\cite{Swanson} are broader. In addition, $\eta\pi$
is a visible channel for both $a_0(1450)$. A week signal was found
in the region of 1200$\sim$1400MeV in the analysis of $\eta \pi^\pm$
(Fig.2(b) of Ref.\cite{Bes6}), which contradicts the large
$a_0(1450)\pi$ channel of the $\eta_H(0^+0^{-+})$ state. We can
exclude the possibility of $X(1870)$ as the $\eta_H(0^+0^{-+})$
hybrid state preliminarily.

The assignment for $X(1870)$ as the $f_H(0^+1^{++})$ hybrid seems
impossible since the theoretical width of $a_1(1260)\pi$ is rather
broad in our results and in the IKP model. If the partial width of
$a_0(980)\pi$ channel is as large as $a_1(1260)\pi$, the total
widths of $X(1870)$ will be much broader than the experimental
value. But the width given by the SS flux tube decay model for the
$f_H(0^+1^{++})$ hybrid is much small. So the possibility of
$X(1870)$ as a $f_H(0^+1^{++})$ hybrid can not be excluded. We
suggest to detect the decay channel of $a_1(1260)\pi$ because this
channel is forbidden for the $\eta_H(0^+2^{-+})$ state in the IKP
flux tube decay model and very small in the SS flux tube decay model
(see Table\ref{tableVIII}). Then the channel of $a_1(1260)\pi$ can
discriminate the state $f_H(0^+1^{++})$ and $\eta_H(0^+2^{-+})$ for
$X(1870)$.

Finally, if $\eta_2(1870)$ is the $\eta_2(2^1D_2)$ state, its decay
width is predicted about 100MeV which is much smaller than the
experiments. However, the difference can be explained by the remedy
of mixing effect. If $X(1870)$ and $\eta_2(1870)$ have the same
quantum numbers, $0^+2^{-+}$, they should mix with each other with a
visible mixing angle. Then the interference enhancement will enlarge
the width of $\eta_2(1870)$. The broad decay width of $\eta_2(1870)$
could be explained naturally. On the other hand, $\eta_2(1870)$ has
been observed in the channel of $a_0(980)\pi$. However, this channel
seems much small if $\eta_2(1870)$ is a pure $2^1D_2$ $n\bar{n}$
meson. The mixing effect will also enlarge this partial width. Here,
we don't plan to discuss the mixing of $X(1870)$ and $\eta_2(1870)$
further for the complex mechanism.

\section{ DISCUSSIONS AND CONCLUSIONS}
A isoscalar resonant structure of $X(1870)$ was observed by BESIII
in the channels $J/\psi\rightarrow\omega
X(1870)\rightarrow\omega\eta\pi^+\pi^-$ recently. Although the mass
of $X(1870)$ is consistent with the $\eta_2(1870)$, the production,
decay width and decay properties are much different. In this paper,
the mass spectrum and strong decays of the $X(1870)$ and
$\eta_2(1870)$ are analyzed.

Firstly, the mass spectrum are studied in the GI potential model and
the RTs framework. In the GI potential model, both $X(1870)$ and
$\eta_2(1870)$ could be the $\eta'_2(1^1D_2)$, $f'_1(2^3P_1)$ and
$\eta_2(2^1D_2)$ states. In RTs, the possible assignments are the
$\eta(3^1S_0)$, $f'_1(2^3P_1)$ and $\eta_2(2^1D_2)$ states. For the
mass spectrum, they are also good hybrid candidates since the masses
overlap the predictions given by different models (see
Table\ref{tableIV}).

Secondly, the processes of a $n\bar{n}$ quarkonium or a $n\bar{n}g$
hybrid meson decaying into the ``$\mathcal {S} + P$'' mesons are
studied under the SU(3)$_f$ symmetry and the diquark-antidiquark
description of the $\mathcal {S}$ mesons. We assumed the processes
obey the OZI rule. We find that the channels of $a_0\pi$,
$\sigma\eta$ and $f_0\eta$ are the dominant when a $n\bar{n}$
quarkonium or a $n\bar{n}g$ hybrid meson decays primarily through
this kind of processes. This result can explain why $X(1870)$ has
been first observed in the $\eta\pi\pi$ channel.

Thirdly, the two-body strong decay of $X(1870)$ is computed in the
$^3P_0$ model. As the $\eta_2(2^1D_2)$ quarkonium, the predicted
width of $X(1870)$ looks much larger than the observations. The
broad resonance, $\eta_2(1870)$, can be a natural candidate for the
$2^1D_2$ $n\bar{n}$ meson. There, we fix the creation strength,
$\gamma$, in two kinds of processes: \ding
{172}.$n\bar{n}\rightarrow n\bar{n}+n\bar{n}$; \ding {173}.
$n\bar{n}\rightarrow n\bar{s}+s\bar{n}$ and $s\bar{s}\rightarrow
n\bar{s}+s\bar{n}$. The functions of creation vertex are determined
as $\gamma(p) = 3.0 + 25\exp(-4 p^2)$ and $\gamma(p) = 1.8 +
4\exp(-10 p^2)$ respectively. Meanwhile, the SHO wave function
scale, $\beta$s, are obtained by the GI potential model.

We have evaluated the magnitude of the partial widths of ``$\mathcal
{S} + P$'' channels by the ratio, $\frac{\Gamma(X(1870) \rightarrow
a_0(980)\pi)}{\Gamma(\eta(1295) \rightarrow a_0(980)\pi)}$, under a
rather crude assumption that $\eta(1295)/X(1870) \rightarrow
a_0(980)\pi$ through a virtual intermediate $1^3P_0$ $q\bar{q}$
meson (see Fig.\ref{fig7}). Then the uncertainties of the coupling
vertex for $1^3P_0(q\bar{q})\rightarrow a_0(980)$ are assumed to be
canceled in the ratio. The total widths of ``$\mathcal {S} + P$''
are evaluated no more than 12.6MeV and $\Gamma(X(1870) \rightarrow
a_0(980)\pi) \leq3.8$MeV. Since $X(1870)$ primarily decay via
$a_0(980)\pi$, it also indicated that the $X(1870)$ can't be
interpreted as the $2^1D_2$ $n\bar{n}$ state.

We also study the $X(1870)$ as a hybrid state in the flux tube
model. Our results agree well with most of predictions given by
Ref.~\cite{Swanson}. $X(1870)$ looks most like the
$\eta_H(0^+2^{-+})$ state for the narrow predicted width, which is
consistent with the experiments. But we can't exclude the
possibility of $0^+1^{++}$. A precise measurement of $a_1(1260)\pi$
is suggested to pin down this uncertainty.

Finally, some important arguments and useful suggestions are given
as follows. \ding {192}.If $\eta_2(1870)$ is the $\eta_2(2^1D_2)$
state, the broad $\pi_2(1880)$ should be isovector partner of
$\eta_2(1870)$. $\pi_2(1880)$ has been interpreted as the
conventional $2^1D_2$ $q\bar{q}$ meson in Ref.~\cite{Li4}. In deed,
the decay channel of $\omega\rho$ is large enough for
$\pi_2(1880)$~\cite{E852}. This observation disfavors the
$\pi_2(1880)$ as a $2^{-+}$ hybrid candidate for the selection rule
that a hybrid meson decaying into two S-wave mesons is strongly
suppressed~\cite{page}. \ding {193}.If $X(1870)$ is a hybrid meson,
we suggest to search its isospin partner in the decay channels of
$J/\psi\rightarrow\rho f_0(980)\pi$ and $J/\psi\rightarrow\rho
b_1(1235)\pi$, which are accessible at BESIII, Belle and BABAR
Collaborations. The decay channel of $b_1(1235)\pi$ is forbidden for
the $\pi_2(2^1D_2)$ quarkonium due to the ``spin selection
rule''~\cite{Burns,flux1}. We also suggest to search the
$\eta_2(1870)$ in the decay channels of $J/\psi\rightarrow\gamma
\rho\rho$ and $J/\psi\rightarrow\gamma\omega\omega$ since these
channels are forbidden for the hybrid production.

\begin{acknowledgments}
Bing Chen thanks Jun-Long Tian and D.V. Bugg for very helpful
discussions. This work is supported by the Key Program of the He'nan
Educational Committee of China (No.13A140014), the National Natural
Science Foundation of China under grant No. 11305003, No. 11075102,
No. 11005003, and U1204115, the Innovation Program of Shanghai
Municipal Education Commission under grant No. 13ZZ066, and the
Program of He'nan Technology Department (No. 11147201).
\end{acknowledgments}

\appendix

\section{The expressions of amplitudes}

We have omitted a exponential factor in following decay amplitudes
$\mathcal {M}_{LS}$ for compactness,

\begin{equation}
\exp(-\frac{2\lambda \mu-\nu^2}{4\mu}p^2).
\end{equation}
where we defined
\begin{eqnarray*}
\mu=\frac{1}{2}(\frac{1}{\beta_A^2}+\frac{1}{\beta_B^2}+\frac{1}{\beta_C^2});\hspace
{0.12cm}\nu=\frac{m_1}{(m+m_1)\beta_B^2}+\frac{m_2}{(m+m_2)\beta_C^2}.
\end{eqnarray*}
and
\begin{eqnarray*}
\lambda=\frac{m_1^2}{(m+m_1)^2\beta_B^2}+\frac{m_2^2}{(m+m_2)^2\beta_C^2};\hspace
{0.36cm} \eta=\frac{m_1}{m+m_1}.
\end{eqnarray*}

For $1^3S_1\rightarrow 1^1S_0 + 1^1S_0$,
\begin{equation}
\mathcal
{M}_{10}=\frac{2\mu-\nu}{8\sqrt{3}\pi^{5/4}\mu^{5/2}(\beta_A\beta_B\beta_C)^{3/2}}p
\end{equation}

For $1^3P_2\rightarrow 1^1S_0 + 1^1S_0$,
\begin{equation}
\mathcal
{M}_{20}=\frac{2\mu\beta_B^{3/2}-(p^2\nu^2+2\mu(1-p^2\nu))\beta_C^{3/2}}{8\sqrt{15}\pi^{5/4}\mu^{7/2}\beta_A^{5/2}\beta_B^{3/2}\beta_C^{3}}
\end{equation}

For $1^3P_2\rightarrow 1^3S_1 + 1^1S_0$, $\mathcal
{M}_{21}=-\sqrt{3/2}\mathcal {M}_{20}$.

For $1^3P_1\rightarrow 1^3S_1 + 1^1S_0$,
\begin{equation}
\mathcal
{M}_{01}=\frac{4\mu\beta_B^{3/2}+(p^2\nu^2+2\mu(1-p^2\nu))\beta_C^{3/2}}{24\pi^{5/4}\mu^{7/2}\beta_A^{5/2}\beta_B^{3/2}\beta_C^{3}}
\end{equation}
\begin{equation}
\mathcal
{M}_{21}=\frac{(2\mu-\nu)\nu}{24\sqrt{2}\pi^{5/4}\mu^{7/2}\beta_A^{5/2}\beta_B^{3/2}\beta_C^{3}}p^2
\end{equation}

For $1^3D_3\rightarrow 1^1S_0 + 1^1S_0$,
\begin{equation}
\mathcal
{M}_{30}=-\frac{2\mu-\nu}{16\sqrt{35}\pi^{5/4}\mu^{9/2}\beta_A^{7/2}\beta_B^{3/2}\beta_C^{3/2}}\nu^2p^3
\end{equation}

\begin{widetext}
For $2^1S_0\rightarrow 1^3P_0 + 1^1S_0$,
\begin{equation}
\begin{aligned}
\mathcal
{M}_{11}=&\frac{1}{96\pi^{5/4}\mu^{11/2}\beta_A^{7/2}\beta_B^{5/2}\beta_C^{3/2}}\\
&\times(-24p^2 \eta \mu^3 - p^4 \nu^4 + 2 p^2 \mu \nu^2 (-10 + p^2 (1 + \eta)\nu) -4 \mu^2 (15 - 5 p^2 (1 + \eta) \nu + p^4 \eta \nu^2)\\
&+6 \mu^2 (4 p^2 \eta \mu^2 + p^2 \nu^2 -2 \mu (-3 + p^2 (1 +
\eta)\nu))\beta_A^2)
\end{aligned}
\end{equation}

For $2^1D_2\rightarrow 1^3S_1 + 1^1S_0$,
\begin{equation}
\mathcal {M}_{11}=\frac{-p^4\nu^4+2p^2\nu^2\mu(\nu
p^2-14)+28\mu^2(\nu p^2-5)+14\mu^2(p^2\nu^2-2(\nu
p^2-5)\mu\beta_A^2)}{160\sqrt{21}\pi^{5/4}\mu^{13/2}\beta_A^{11/2}\beta_B^{3/2}\beta_C^{3/2}}\nu
p
\end{equation}
\begin{equation}
\mathcal {M}_{31}=\frac{-28\mu^2+\nu^3p^2-2\mu\nu(\nu
p^2-9)+14(2\mu-\nu)\mu^2\beta_A^2}{160\sqrt{14}\pi^{5/4}\mu^{13/2}\beta_A^{11/2}\beta_B^{3/2}\beta_C^{3/2}}\nu^2p^3
\end{equation}

For $2^1D_2\rightarrow 1^3S_1 + 1^3S_1$, $\mathcal
{M}'_{11}=\sqrt{2}\mathcal {M}_{11}$ and $\mathcal
{M}'_{31}=\sqrt{2}\mathcal {M}_{31}$.

For $2^1D_2\rightarrow 1^3P_0 + 1^1S_0$
\begin{equation}
\begin{aligned}
\mathcal {M}_{20}=&-\frac{1}{192\sqrt{35}\pi^{5/4}\mu^{15/2}\beta_A^{11/2}\beta_B^{5/2}\beta_C^{3/2}}\\
&\times( p^4 \nu^5 - 2 p^2 \mu \nu^3 (-18 + p^2 (1 + \eta) \nu) +
  4 \mu^2 \nu (63 - 11 p^2 (1 + \eta) \nu +
     p^4 \eta \nu^2) +
  56 \mu^3 (-2 + \eta (-2 + p^2 \nu)) \\
&- 14 \mu^2 (p^2 \nu^3 -
     2 \mu \nu (-7 + p^2 (1 + \eta) \nu) +
     4 \mu^2 (-2 + \eta (-2 + p^2\nu)))\beta_A^{2})\nu p^2.
\end{aligned}
\end{equation}

For $2^1D_2\rightarrow 1^3P_1 + 1^1S_0$
\begin{equation}
\begin{aligned}
\mathcal {M}_{21}=-\frac{p^2 \nu^2 - 14 \mu^2 \beta^2_A+
14\mu}{16\sqrt{35}\pi^{5/4}\mu^{11/2}\beta_A^{11/2}\beta_B^{5/2}\beta_C^{3/2}}(\eta
- 1)\nu p^2 .
\end{aligned}
\end{equation}

For $2^1D_2\rightarrow 1^3P_2 + 1^1S_0$
\begin{equation}
\begin{aligned}
\mathcal {M}_{02}=&\frac{1}{480\sqrt{14}\pi^{5/4}\mu^{15/2}\beta_A^{11/2}\beta_B^{5/2}\beta_C^{3/2}}\\
&\times(p^6 \mu^6 - 2 p^4 \mu \nu^4 ( p^2 (1 + \eta) \nu -21) +
  4 p^2 \mu^2 \nu^2 (105 - 14 p^2 (1 + \eta) \nu +
     p^4 \eta \nu^2) +
  56 \mu^3 (15 - 5 p^2 (1 + \eta) \nu \\
&+ p^4 \eta \nu^2) -
  14 \mu^2 (p^4 \nu^4 -
     2 p^2 \mu \nu^2 (-10 + p^2 (1 + \eta) \nu) +
     4 \mu^2 (15 - 5 p^2 (1 + \eta) \nu + p^4 \eta \nu^2))\beta_A^{2}).
\end{aligned}
\end{equation}

\begin{equation}
\begin{aligned}
\mathcal {M}_{22}=&-\frac{1}{672\sqrt{5}\pi^{5/4}\mu^{15/2}\beta_A^{15/2}\beta_B^{5/2}\beta_C^{3/2}}\\
&\times(p^4 \nu^5 - 2 p^2 \mu \nu^3 (-18 + p^2 (1 + \eta) \nu) +
 2 \mu^2 \nu (126 - 25 p^2 (1 + \eta) \nu +
    2 p^4 \eta \nu^2) +
 28 \mu^3 (-7 + \eta (-7 + 2 p^2 \nu)) \\
&-14 (\mu^2) (p^2 \nu^3 -
    2 \mu \nu (-7 + p^2 (1 + \eta) \nu) +
    2 \mu^2 (-7 + \eta (-7 + 2 p^2 \nu))) \beta_A^{2})\nu p^2 .
\end{aligned}
\end{equation}

\begin{equation}
\begin{aligned}
\mathcal {M}_{42}=&-\frac{1}{1120\pi^{5/4}\mu^{15/2}\beta_A^{11/2}\beta_B^{5/2}\beta_C^{3/2}}\\
&\times(\nu (  p^2 \nu^3-36 \mu^2 + 2 \mu \nu (11 - p^2 \nu)) +
 2 \eta \mu (28 \mu^2 - p^2 \nu^3 +
    2\mu\nu (p^2 \nu-9)) -
 14 \mu^2 (2 \mu - \nu) (2 \eta \mu - \nu) \beta_A^{2})\nu^2 p^4 .
\end{aligned}
\end{equation}

For $2^1D_2\rightarrow 2^3P_1 + 1^1S_0$
\begin{equation}
\begin{aligned}
\mathcal {M}_{21}=&\frac{1}{160\sqrt{14}\pi^{5/4}\mu^{15/2}\beta_A^{11/2}\beta_B^{9/2}\beta_C^{3/2}}\\
&\times(-112 \eta \mu^3 + 252 \mu^2 \nu +
  56 p^2 \eta^2 \mu^3 \nu - 80 p^2 \eta \mu^2 \nu^2 +
  36 p^2 \mu \nu^3 + 4 p^4 \eta^2 \mu^2 \nu^3 -
  4 p^4 \eta \mu \nu^4 + p^4 \nu^5 \\
&-10 \mu^2 \nu (14 \mu + p^2 \nu^2) \beta_B^2 - 14 \mu^2
\beta_A^2(14 \mu \nu + 4 p^2 \eta^2 \mu^2 \nu + p^2 \nu^3 - 4 \eta
\mu (2 \mu + p^2 \nu^2) - 10 \mu^2 \nu\beta_B^2))(\eta-1)p^2 .
\end{aligned}
\end{equation}

\end{widetext}

$m_1$ and $m_2$ are the masses of quarks in the decaying meson
\emph{A}. \emph{m} is the mass of the created quark from the vacuum.
For calculating the decay widths, the masses of quarks are taken as:
$m_u$ = $m_d$ = 0.220 GeV, $m_s =$ 0.428 GeV, which are as same as
these in the Section II. The above amplitudes, $\mathcal {M}_{LS}$,
can be reduced further in the approximation of $m_1 = m_2 = m$ and
$\beta_A = \beta_B = \beta_C = \beta$. The reduced $\mathcal
{M}_{LS}$ are consistent with these given by Ref.~\cite{Barnes}
except for an unimportant factor, $-2^{9/2}$, since this factor can
be absorbed into the coefficient $\gamma$.

\section{Hybrid decay in the flux tube model}

The flux tube model was motivated by the strong coupling expansion
of the lattice QCD. In this model, decay occurs when the flux-tube
breaks at any point along its length, with a $q\bar{q}$ pair
production in a relative $J^{PC} = 0^{++}$ state. It is similar to
the $^3P_0$ decay model but with an essential difference. The flux
tube model extend the nonrelativistic constituent quark model to
include gluonic degrees of freedom in a very simple and intuitive
way, where the gluonic field is regarded as tubes of color flux.
Then it can be extended to the hybrid research. When the hybrid
mesons are assumed to be narrow, and the threshold effects aren't
taken into account, the partial decay width
$\Gamma_{LS}(H\rightarrow BC)$ is given by the flux model
as~\cite{flux1}

\begin{equation}
\Gamma_{LS}(H\rightarrow
BC)=\frac{p}{(2J_A+1)\pi}\frac{\tilde{M}_B\tilde{M}_C}{\tilde{M}_A}|\mathcal
{M}_{LS}(H\rightarrow BC)|^2
\end{equation}
where $\tilde{M}_A$, $\tilde{M}_B$, $\tilde{M}_C$ are the
``mock-meson'' masses of A, B, C~\cite{Kokoski}. When a hybrid meson
decay into \emph{P}-wave and pseudoscalar mesons, the partial wave
amplitude $\mathcal {M}_{L}(H\rightarrow BC)$ (with $S=S_B$) is
given as the following form

\begin{widetext}

\begin{equation}
\mathcal {M}_L(H\rightarrow
BC)=\langle\phi_B\phi_C|\phi_A\phi_0\rangle(\frac{a\tilde{c}}{9\sqrt{3}}\frac{1}{2}A^0_{00}\sqrt{\frac{fb}{\pi}})\frac{\kappa\sqrt{b}}{(1+fb/(2\tilde{\beta}^2))^2}\sqrt{\frac{2\pi}{3\Gamma(3/2+\delta)}}\frac{\beta_A^{3/2+\delta}}{\tilde{\beta}}\tilde{\mathcal
{M}}_L(H\rightarrow BC)
\end{equation}

\end{widetext}

The flavor matrix element $\langle\phi_B\phi_C|\phi_A\phi_0\rangle$
have been discussed before. $\tilde{\mathcal {M}}_L(H\rightarrow
BC)$ are listed in Table \ref{tableIX} for the states of
$\eta_H(0^+0^{-+})$, $f_H(0^+1^{++})$ and $\eta_H(0^+2^{-+})$.

\begin{table}
\begin{ruledtabular}
\begin{tabular}{llll}
   \emph{B}     & \emph{H}($0^+0^{-+}$) &  \emph{H}($0^+1^{++}$)                   &  \emph{H}($0^+2^{-+}$)  \\
\hline
$0^{++}$&   $+\sqrt{2}\mathcal {M}_S/3$ &  $-\sqrt{2}\mathcal {M}_{P_2}/\sqrt{3}$  &  $+\mathcal {M}_D/3$\\
$1^{++}$&   $-$                         &  $-\mathcal {M}_{P_1}/\sqrt{2}$          &  - \\
$2^{++}$&   $+\mathcal {M}_D/3$         &  $-\mathcal {M}_{P_4}/\sqrt{30}$         & $-\sqrt{5}\mathcal {M}_S/\sqrt{18}$  \\
        &                               &  $+\mathcal {M}_F/\sqrt{5}$              & $-\sqrt{7}\mathcal {M}_D/3$  \\
\end{tabular}
\end{ruledtabular}
\caption{\label{tableIX}Partial wave amplitudes $\tilde{\mathcal
{M}}_L(H\rightarrow BC)$ for an initial hybrid \emph{H} decaying
into a \emph{P}-wave and pseudoscalar mesons. }
\end{table}

Here the $\mathcal {M}_S$, $\mathcal {M}_D$, $\mathcal {M}_{P_i}$
and $\mathcal {M}_F$ are defined as $\mathcal {M}_S = -(3\tilde{h}_0
- \tilde{g}_1+4\tilde{h}_2)$, $\mathcal {M}_D = (\tilde{g}_1+5
\tilde{h}_2)$, $\mathcal {M}_{P_1} =
-i(2\tilde{g}_0+3\tilde{h}_1-\tilde{g}_2)$, $\mathcal {M}_{P_2} =
-i(\tilde{g}_0+\tilde{g}_2)$, $\mathcal {M}_{P_4} =
-i(10\tilde{g}_0+9\tilde{h}_1+\tilde{g}_2)$ and $\mathcal {M}_F =
-3i(\tilde{g}_2+\tilde{h}_3)$. The analytical expressions of
$\tilde{g}_i$ and $\tilde{h}_i$ are given as

\begin{widetext}
\begin{equation}
\tilde{g}_n =
2^{3+\delta}\frac{M^nm}{(M+m)^{n+1}}(2\beta_A^2+\tilde{\beta}^2)^{-\frac{n+\delta+3}{2}}\Gamma(\frac{n+\delta+3}{2})_1F_1[\frac{n+\delta+3}{2},
n+1,
-(\frac{M}{M+m})^2\frac{p^2}{2\beta_A^2+\tilde{\beta}^2}]p^{n+1}
\end{equation}
\begin{equation}
\tilde{h}_n =
2^{3+\delta}\tilde{\beta}^2(\frac{M}{M+m})^n(2\beta_A^2+\tilde{\beta}^2)^{-\frac{n+\delta+4}{2}}\Gamma(\frac{n+\delta+4}{2})_1F_1[\frac{n+\delta+4}{2},
n+1, -(\frac{M}{M+m})^2\frac{p^2}{2\beta_A^2+\tilde{\beta}^2}]p^n
\end{equation}
\end{widetext}
where $_1F_1[\cdots]$ are the confluent hypergeometric functions.
Here we don't take account of the decay channels of $H\rightarrow
2S+1S$ because they are forbidden by the conservation laws, or the
``spin selection rule'', or the phase space, \emph{e.g.}, the decay
channel of $\pi(1300) + \pi$ is forbidden for the $f_H(0^+1^{++})$
state by the ``spin selection rule''. In this work, we choose to
follow the Refs.~\cite{flux1} and take the combination
$(a\tilde{c}/9\sqrt{3})\frac{1}{2}A^0_{00}\sqrt{\frac{fb}{\pi}}$ as
0.64 which was fixed by the conventional mesons~\cite{Kokoski}, $M =
m = m_{u,d} = 330$MeV, $\tilde{M}^{I=0}_B$ $= \tilde{M}^{I=1}_B =
1250$MeV, $\tilde{M}^{I=0}_C = 720$MeV, $\tilde{M}^{I=1}_C =
850$MeV, $\beta_A =$0.27GeV, $\delta =$0.62, $b =$0.18GeV$^2$ and
$\kappa =$0.9. Final states containing $\pi$ have $\tilde{\beta}
=$0.36GeV, otherwise $\tilde{\beta} =$0.40GeV.

\newpage %Just because of unusual number of tables stacked at end

\end{document}